\magnification=\magstephalf \font\t=cmcsc10 at 13 pt
 \font\n=cmcsc10 \font\foot=cmr9 \font\foots=cmsl9
\font\abs=cmr8 \font\babs=cmbx8  \centerline{\t The
singular points of Einstein's universe \footnote{\dag}{\foot{Le
Journal de Physique et Le Radium {\bf 23}, 43 (1923).}}}\bigskip
\centerline{\n by Marcel Brillouin}\bigskip \vbox to 0.6 cm {}
\centerline{(translation by S. Antoci\footnote{$^*$}{\foot Dipartimento di Fisica
``A. Volta'', Universit\`a di Pavia, Via Bassi 6 - 27100 Pavia
(Italy).})}\medskip {\babs } {\abs }\par \vbox to 0.6 cm
{} 1. Einstein's four-dimensional Universe is determined by the
ten $g_{\mu\nu}$ of its $ds^2$. In order to determine them it is
not sufficient to know the six independent partial differential
equations that they must fulfil; one needs also to know the conditions
at the boundaries of the Universe, which are necessary for
specialising the integrals in view of given problems. These
boundary conditions are of two sorts. One deals with the far away
state of the Universe, completely outside the region where we wish
to study the events; it is the one whose choice, still in
dispute, is translated into this question: is the Universe
infinite? Is it finite, although without limit? I do not bother with
this here. The other one deals with the singular lines that
correspond to what, from the experimental viewpoint, we call the
attractive masses. In Newtonian gravitation the material point of
mass $m$ corresponds to the point of the Euclidean space where
the integral of Laplace's equation becomes infinite like $m/r$,
where $r$ is (in the neighbourhood of this point) the distance
from the material point to the point where one studies the
Newtonian potential. It is this kind of singularities,
characteristic of matter, that I come to consider.\par We first remark
that, in the present state of our experimental knowledge, nothing
entitles us to suppose that singular points (in four dimensions)
may exist in the Universe. From the analytical viewpoint, this
impossibility is evidently connected with the distinction that
exists between one of the variables and the remaining three, which allows
for the sign changes of $ds^2$, like in acoustics. It would be
interesting to give precision to this remark.\par\smallskip
2. Let us consider a static, permanent state, {\it i.e.} a state in
which the $g_{\mu\nu}$ depend only on three of the variables,
$x_1$, $x_2$, $x_3$, and are independent of that $x_4$ whose
$g_{44}$ in essentially positive. The simplest singular line of
Universe is the one which, in the section with the three dimensions $x_1$,
$x_2$, $x_3$ that we call {\it space}, corresponds to an {\it
isotropic singular point}.\par For a Universe that contains only
one line of this kind Schwarzschild has integrated, in 1916, the
differential equations that rule the $g_{\mu\nu}$, and he obtained a
$ds^2$ that, by changing Schwarzschild's notations, I write under
the form
$$ds^2=\gamma c^2d\tau^2-{1\over\gamma}dR^2
-(R+2m)^2(d\theta^2+sin^2\theta d\varphi^2),$$
$$\gamma={R\over\ {R+2m}},$$
$c$, velocity of light (universal constant);\par\noindent
$\theta$, $\varphi$, the spherical polar angles in space;\par\noindent
$R$, a length such that the sphere,
whose centre is at the singular point in the (non Euclidean) space
$R$, $\theta$, $\varphi$, and which has $R$ for co-ordinate
radius, has a total surface equal to $4\pi(R+2m)^2$, and $2\pi(R+2m)$
as circumference of the great circle.\par The function $\gamma$ is
positive, and equal to 1 at a large distance; $m$ is a positive
constant. If $R$ is not zero, $\gamma$ is finite and
nonvanishing; there is no singularity either of $g_{\mu\nu}$ or of
$ds^2$.\par But, if $R$ is zero, $\gamma$ is zero, and the
coefficient of $dR^2$ is infinite: there is a pointlike singularity
in this point in space, and a line singularity in the
Universe.\par\smallskip
3. One may wonder whether this singularity {\it limits} the Universe,
and one must stop at $R=0$ or, on the contrary, it only {\it
traverses} the Universe, which shall continue on the other side,
for $R<0$. In the discussions at the ``Coll\`ege de France'', in
particular in the ones held during the Easter of 1922, it has
been generally argued as if $R=0$ would mean a {\it catastrophic}
region that one needs to cross in order to attain the true,
singular limit, that one only reaches when $\gamma$ is infinite,
with $R=-2m$. {\it In my opinion, it is the first singularity,
reached when $R=0$, $\gamma=0$ $(m>0)$, the one that limits the
Universe} and that must not be crossed [{\it C. R.}, {\bf 175},
(November 27th 1922)].\par
    The reason for this is peremptory, although up to now I have
neglected to put it in evidence: {\it for $R<0$, $\gamma<0$, in
no way the $ds^2$ any longer corresponds to the problem one aimed
at dealing with}.\par
In order to see it clearly, let us put anew the letters
$x_1...x_4$ whose physical meaning shall not be suggested by
old habits. Schwarzschild's $ds^2$ corresponds to the following
analytical problem: the $g_{\mu\nu}$ depend only on the single
variable $x_1$; two more variables, $\theta$ and $\varphi$, enter
in the manner that corresponds to the spatial isotropy around one
point, and one has:
$$ds^2=\gamma c^2dx_4^2-{1\over\gamma}dx_1^2
-(x_1+2m)^2(d\theta^2+sin^2\theta d\varphi^2),$$
$$\gamma={x_1\over\ {x_1+2m}}~~~m>0.$$
The term $\gamma$ is positive either when
$$x_1>0~~and~~x_1+2m>0$$
or when
$$x_1<0~~and~~x_1+2m<0;$$
one has truly solved the proposed problem, $x_1$ is a spatial
variable and $x_4$ is a time variable.\par\smallskip
4. {\it If $\gamma$ is instead negative, $-2m<x_1<0$, the characters
of length and of duration are exchanged between $x_1$ and $x_4$};
in fact now the term $-(1/\gamma)dx_1^2$ is positive, while the
term $\gamma c^2dx_4^2$ is negative.\par
Let us make this character visible by substituting the notation
$t$ for $-x_1$:
$$ds^2={{2m-t}\over t}dt^2-{t \over {2m-t}}dx_4^2
-(2m-t)^2(d\theta^2+sin^2\theta d\varphi^2).$$

This is a $ds^2$ which has no longer any relation with the static
problem that one aimed at dealing with. {\it The $ds^2$ for $x_1<0$
does not continue the one that is appropriate for $x_1>0$}. This
discontinuity is by far sharper than all the ones that have been
encountered up to now in the problems of mathematical physics. The
frontier $x_1=0$, $R=0$, is really an insurmountable one.\par
While discussing Schwarzschild's integration one notices that an
arbitrary factor $C_4$ could have been left in the term with
$dx_4$, and that this factor is taken equal to 1 in order that the
Universe become Euclidean when $x_1$ is infinite. No similar
condition can be imposed in the interval $-2m<x_1<0$; but $C_4$ is
a real constant, and it can only be taken with a positive value.
In fact, if it were negative, the $ds^2$ would no longer have any
meaning that refer to an Einstein's universe.\par\smallskip
5. The conclusion seems to me unescapable: the limit $R=0$ is
insurmountable; it embodies the {\it material} singularity.\par
The distance $r$ from this origin ($R=0$) to a point with
co-ordinate radius $R$, calculated along a radius vector
($\theta=const., \varphi=const.)$ is \footnote{$^{(1)}$}
{{\foots Erratum.} {\foot In the already cited note of
the {\foots C. R.} the coefficient $-m/2$ of the logarithm is
incorrect.}}
$$r=\sqrt{R(R+2m)}+m\ln{{R+m+\sqrt{R(R+2m)}\over m}}.$$

The ratio between the circumference $2\pi(R+2m)$ and the radius is
everywhere larger than $2\pi$; in particular at the origin $(r=0,
R=0)$ this ratio becomes infinite. The circumference of the great
circle of the origin is $4\pi m$; the spherical surface of that
point has the finite value $4\pi(2m)^2$. It is this singularity
that constitutes what physics calls {\it the material point}; it is
the factor $m$ appearing in it that must be called mass.\par
In view of this occurrence, the word {\it material point} is
perhaps ill chosen. In fact, due to the finite extension of the
spherical surface of the point, the variations of $\theta$ and of
$\varphi$ truly displace the extremity of the radius vector (of
zero length) over this surface as it would occur over any other sphere
whose co-ordinate radius $R$ and whose radius $r$ were
nonvanishing.\par
Anyway, since nothing more pointlike can be found in
Einstein's Universe, and since one really needs attaining a
definition of the elementary {\it material test body} which, according
to Einstein, follows a geodesic of the Universe to which it belongs, I will
maintain this {\it abridged nomenclature, material point}, without
forgetting its imperfection.\par\smallskip
{\abs January 1923.}
\end